\documentclass{elsart}

\usepackage{graphicx,tabularx}

\begin{document}
\begin{frontmatter}



\title{Bulk-sensitive Photoemission of Mn$_{5}$Si$_{3}$}
\author[label1]{A. Irizawa},
\author[label1]{A. Yamasaki},
\author[label1]{M. Okazaki},
\author[label1]{S. Kasai},
\author[label1]{A. Sekiyama},
\author[label1]{S. Imada},
\author[label1]{S. Suga},
\author[label2]{E. Kulatov},
\author[label3]{H. Ohta},
\author[label4]{T. Nanba}
\address[label1]{Graduate School of Engineering Science, Osaka University, Toyonaka, Osaka 560-8531, Japan}
\address[label2]{General Physics Institute, Russian Academy of Science, Moscow, Russia}
\address[label3]{Molecular Photoscience Research Center, Kobe University, Rokkodai, Nada, Kobe 657-8501, Japan}
\address[label4]{The Graduate School of Science and Technology, Kobe University, Rokkodai, Nada, Kobe 657-8501, Japan}


\begin{abstract}
We have carried out a bulk-sensitive high-resolution photoemission experiment on Mn$_{5}$Si$_{3}$. The measurements are performed for both core level and valence band states. The Mn core level spectra are deconvoluted into two components corresponding to different crystallographic sites. The asymmetry of each component is of noticeable magnitude. In contrast, the Si 2p spectrum shows a simple Lorentzian shape with low asymmetry. The peaks of the valence band spectrum correspond well to the peak positions predicted by the former band calculation.
\end{abstract}


\end{frontmatter}

\section{Introduction}
Transition metal silicide Mn$_{5}$Si$_{3}$ has been known to show complex 
magnetic phase transitions \cite{rf:1,rf:2,rf:3,rf:4,rf:5}. Its crystal structure is 
D8$_{8}$ and has two inequivalent crystallographic sites; namely, two Mn 
atoms in the 4(d) position (Mn$_{I}$ 4(d)), as well as three Mn atoms and 
three Si atoms in the 6(g) position (Mn$_{II}$ 6(g)) in the formula unit 
cell \cite{rf:6,rf:7}. This compound shows the Curie-Weiss type paramagnetic state 
over 99 K, collinear antiferromagnetic state between 66 and 99 K, and 
non-collinear antiferromagnetic state below 66 K with a complex spin 
structure \cite{rf:4}. Below 66 K, it also shows a field-induced magnetic transition 
which is attributed to its topological frustration of spin alignment in the 
non-collinear antiferromagnetic structure \cite{rf:1,rf:2}. The electric resistivity 
represents a metallic behavior with two kinks corresponding to these phase 
transitions \cite{rf:1,rf:5}. Some conceivable magnetic structures are proposed for 
the collinear and non-collinear magnetic phases from the neutron diffraction 
studies \cite{rf:8,rf:9,rf:10,rf:11,rf:12}. For the non-collinear phase at lower 
temperatures, for example, it is considered that several Mn sites with 
different magnitude and orientation of the magnetic moments are providing 
itinerant electrons. However, their electronic states are still unclear. 
Thus, measurements of core-level and valence band photoemission are very 
important to reveal their complex electronic states.

In this study, we have performed a bulk-sensitive high-resolution 
photoemission spectroscopy of Mn$_{5}$Si$_{3}$ for the paramagnetic state.

\section{Experimental}
Single crystals of Mn$_{5}$Si$_{3}$ were prepared by Czochralski method. 
Clean surfaces were obtained by scraping with a diamond file. Measurements 
were carried out at the beam line BL25SU in SPring-8 by using a SCIENTA-200 
electron analyzer under the base pressure of better than 5$\times $10$^{ - 
10}$ Torr. The excitation photon energy was set to 900 eV for the 
bulk-sensitive experiments. The total energy resolution was set to about 100 
meV for the valence band region and about 200 meV for the core-level 
photoemission. These values are used for the analyses of spectral 
deconvolution as the Gaussian width as performed later. The measurements 
were done at 195 K at which Mn$_{5}$Si$_{3}$ is in the Curie-Weiss type 
paramagnetic state with metallic conductivity \cite{rf:1,rf:4}.

\section{Results and Discussion}
Figures 1(a)-(d) show the core-level photoemission spectra of 
Mn$_{5}$Si$_{3}$. The Si 2p doublet ascribable to the j=3/2 and 1/2 
spin-orbit components is clearly observed as shown in Figure 1(a). The 
spectral shape can be fit by a single component for both j=3/2 and 1/2 
components by adopting the Doniach-Sunjic line shape \cite{rf:13}. The asymmetry 
parameter of $\alpha $=0.017 is employed for both j=3/2 and 1/2 peaks. The 
values of the experimental full width at half maximum (FWHM) are 0.43 eV for 
the j=3/2 peak and 0.42 eV for the j=1/2 peak (see Table 1).

\begin{figure}
\includegraphics[width=6.5cm,clip]{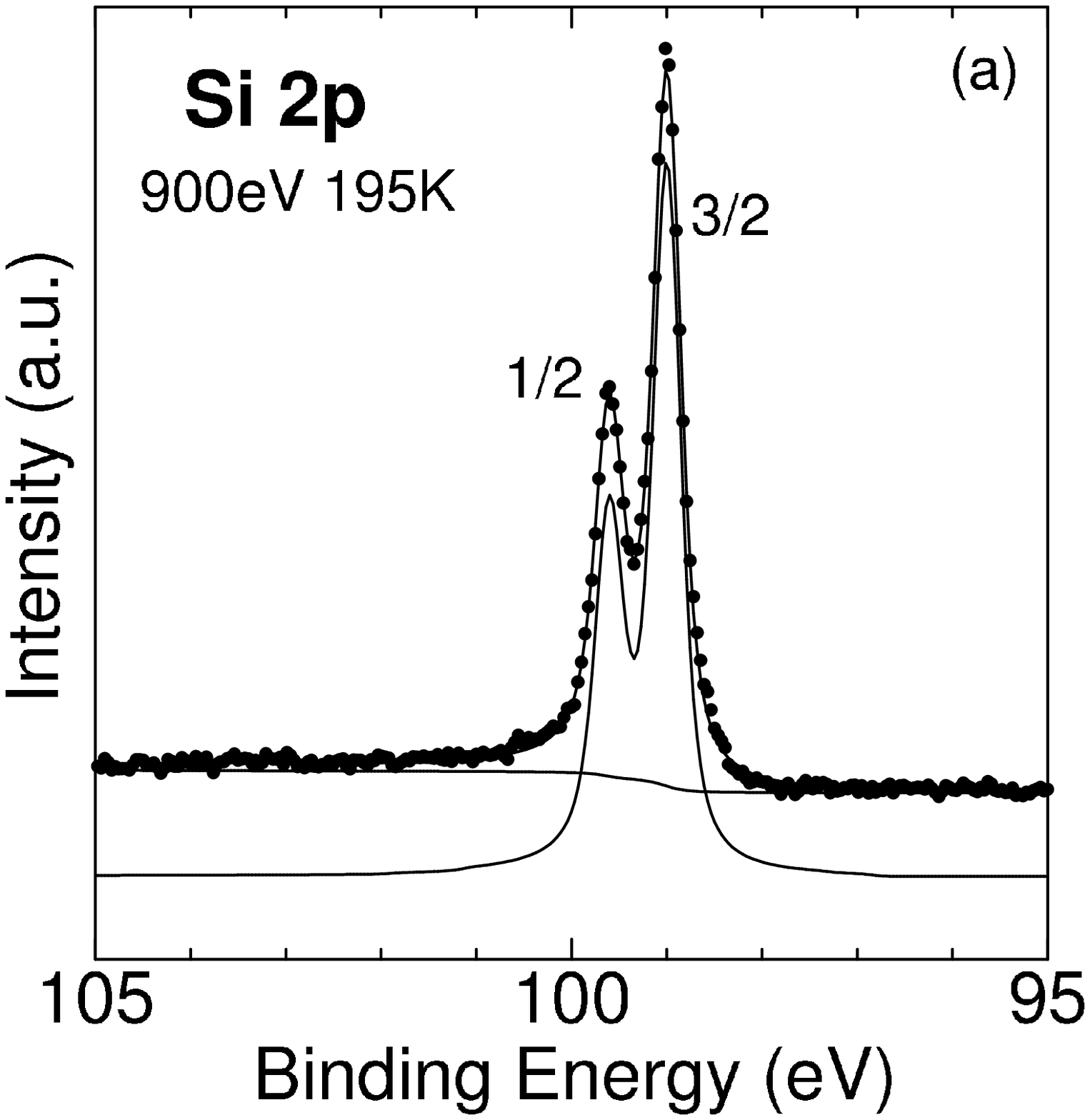}
\includegraphics[width=6.5cm,clip]{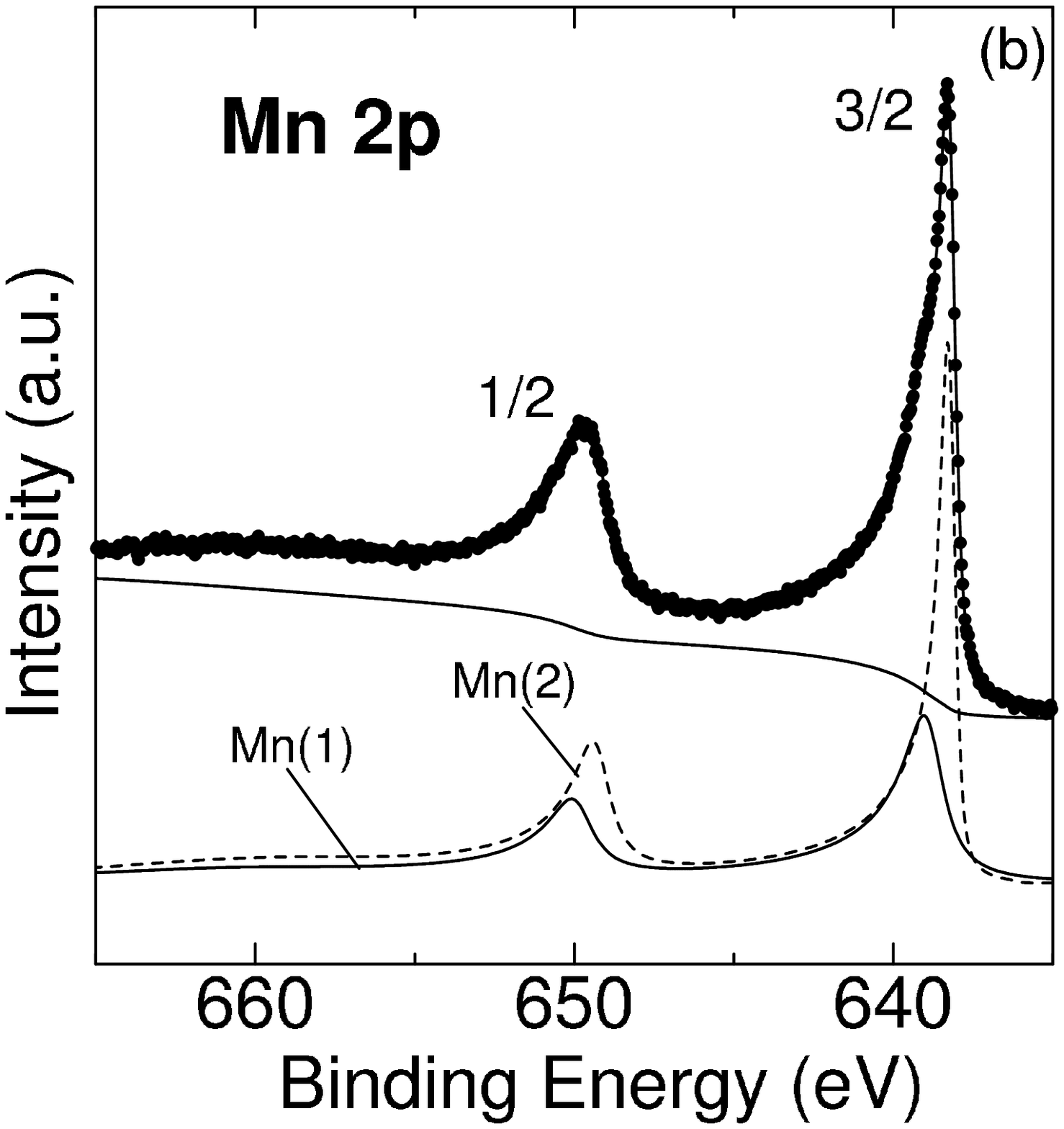}

\includegraphics[width=6.5cm,clip]{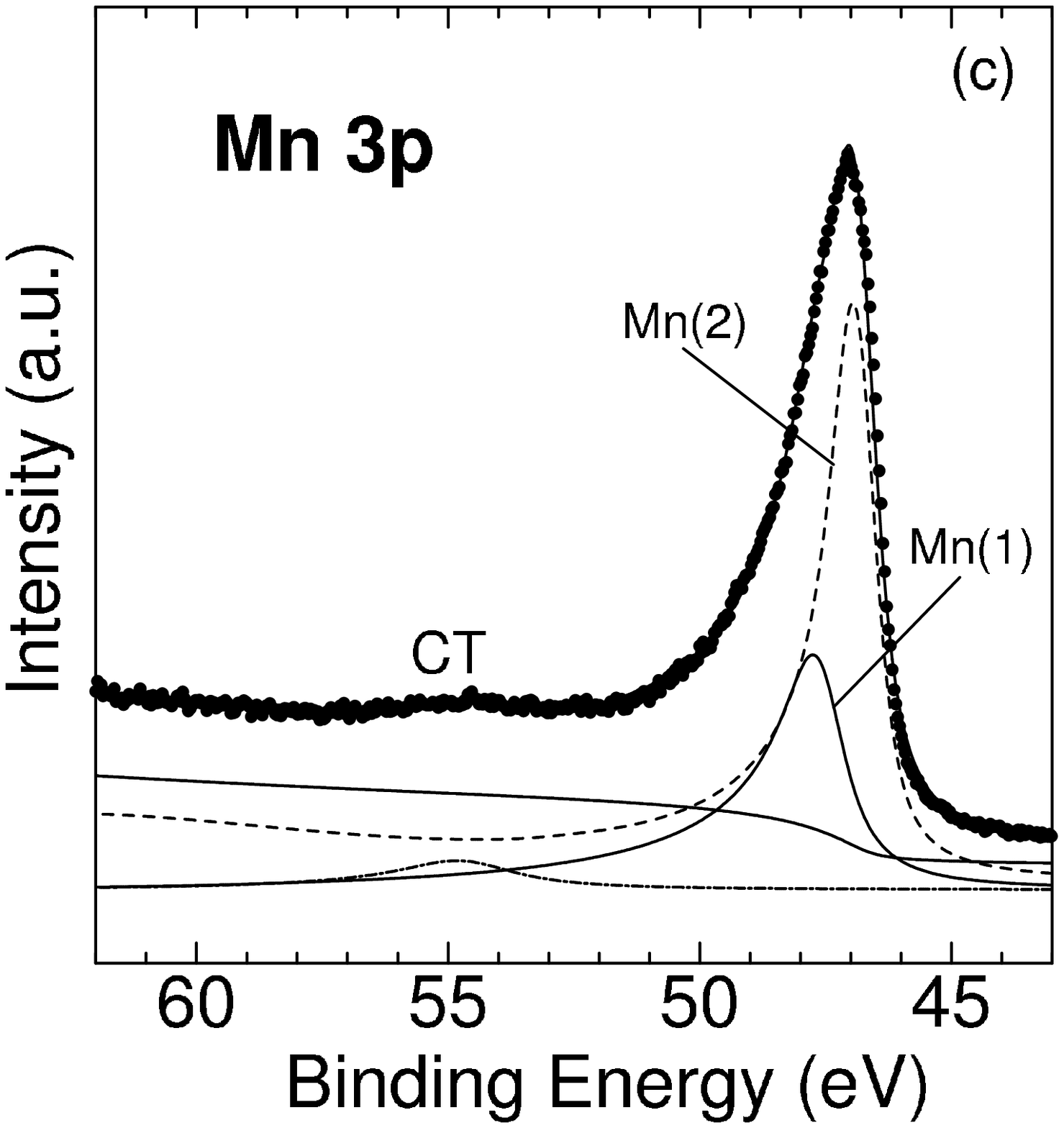}
\includegraphics[width=6.5cm,clip]{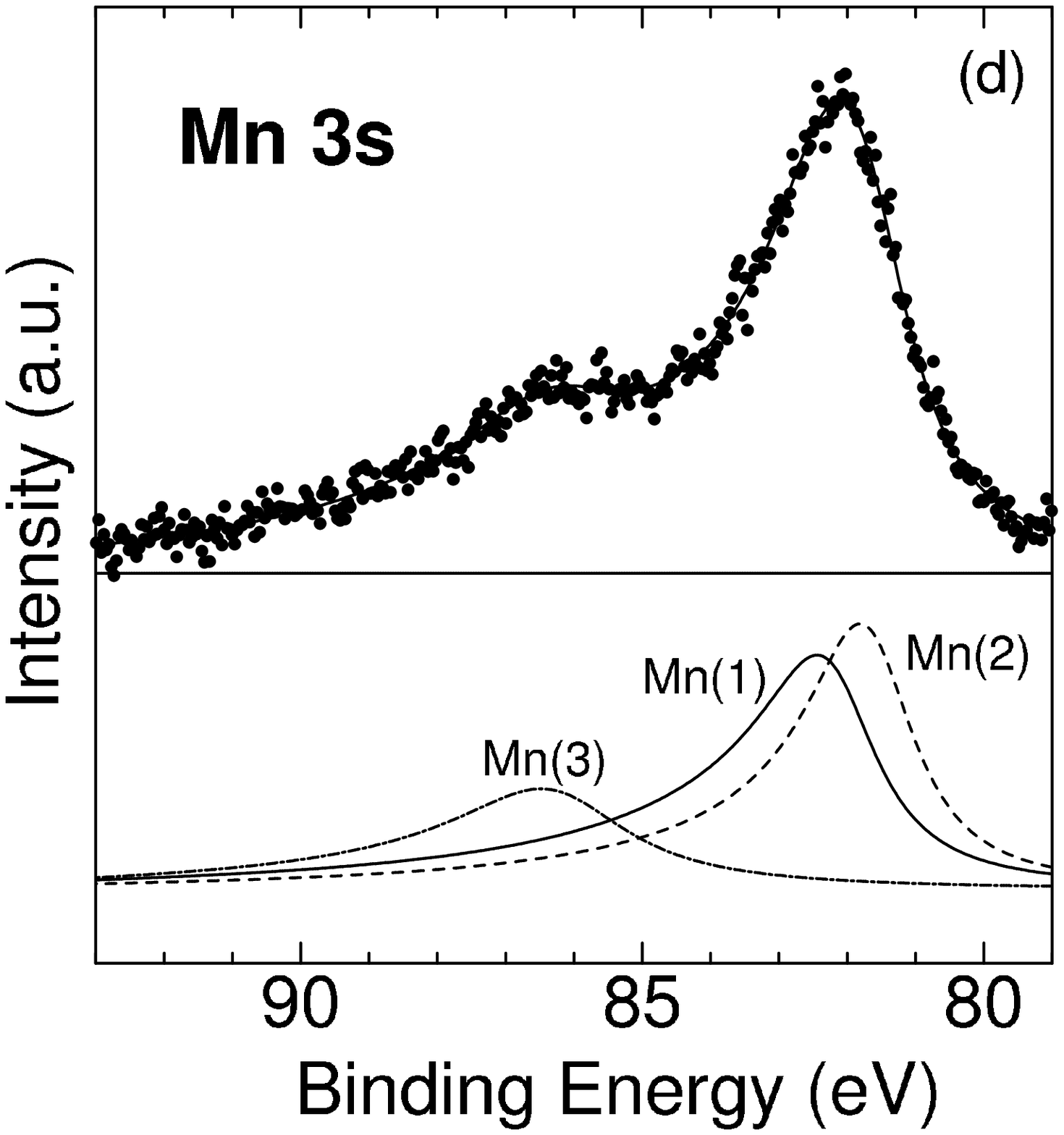}
\caption{Core-level photoemission spectra of Mn$_5$Si$_3$ measured at 195 K for h$\nu$=900 eV; (a) Si 2p, (b) Mn 2p, (c) Mn 3p, and (d) Mn 3s states. Dots are the raw data and the solid curves through them are the results of fittings. The other curves are the respective components and the secondary electron background for the fittings.}
\label{fig:1}
\end{figure}%

In contrast, the Mn core-level spectra show broader peaks with satellites. 
Figure 1(b) shows the Mn 2p spectrum split by the spin-orbital interaction. 
The peaks at E$_{B}$=638 eV and 650 eV correspond to the j=3/2 and j=1/2 
states. The j=3/2 peak cannot be explained by a single asymmetric component. 
It seems to be composed of a weaker peak at higher E$_{B}$ and a stronger 
peak at lower E$_{B}$. Therefore, both j=3/2 and 1/2 peaks are deconvoluted 
into two components as labeled by Mn(1) and Mn(2). The asymmetry parameter 
$\alpha $ for the Mn(2) component is evaluated to be slightly smaller than 
that for the Mn(1) component as shown in Table 1. The intensity ratio of the 
Mn(1) and Mn(2) is roughly 2:3. The energy separation of them is 0.66 eV. 
The values of FWHM of the Mn 2p spectrum are roughly three times larger than 
that of Si 2p spectrum except for the Mn(2) component in the j=3/2 peak of 
the Mn 2p spectrum.

Next, we show the Mn 3p spectrum in Fig. 1(c). A single prominent peak is 
observed at E$_{B}$=47 eV. Due to the 3p-3d multiplet effect, the spin-orbit 
splitting is not directly observed. There appears a small hump near 
E$_{B}$=55 eV in addition to the main peak. This hump may correspond to the 
charge-transfer (CT) satellite of the main peak. It is not clearly 
recognized in other Mn core-level spectra because of its smallness. The 
smallness of this CT satellite intensity implies the dominance of the d$^{n 
+ 1}$\underline {L} final state for the main peak. Namely, the 3p core hole 
is well screened by the ligand electron. This is consistent with the good 
electric conductivity of this compound \cite{rf:1,rf:5}. The main peak at E$_{B}$=47 eV 
can not be fit by a single asymmetric Lorentzian but still be deconvoluted 
into two components of Mn(1) and Mn(2) with small and large intensities 
separated by the binding energy of 0.98 eV (see Table 1). The asymmetry 
$\alpha $ of Mn(1) component is higher than that of Mn(2). The intensity 
ratio of them is again roughly estimated as 2:3. All the values of FWHM in 
the Mn 3p components are three to five times larger than that of the Si 2p 
components.

In Fig. 1(d) is shown the Mn 3s photoemission spectrum. There are two broad 
peaks separated by $\sim $4 eV. At first, the larger peak was tentatively 
deconvoluted into two components of Mn(1) and Mn(2) with the intensity ratio 
of 2:3 as in the case of Mn 2p and 3p spectra, where the FWHM was also 
larger for the Mn(1) component. The smaller peak at E$_{B}\sim $86.5 eV 
can be mostly attributed to the spin-exchange interaction. In 3d transition 
metal compounds, a well separated satellite by spin-exchange interaction can 
only be seen for the 3s orbital. This is because the multiplet effect in the 
3s spectrum is less prominent than in the other core-level spectra and the 
magnitude of the exchange integral between the 3d and each core-level is the 
largest for the 3s level \cite{rf:14,rf:15,rf:16}. Then, we also deconvoluted this small 
peak into two components corresponding to the Mn(1) and Mn(2) components by 
assuming a relative intensity of 2:3. Then, the intensities of both 
components in the small peak were evaluated to be only $\sim $26{\%} of the 
sum of the Mn(1) and Mn(2) components in the main peak.

Judging from the energy separation of $\sim $4 eV between the main peak and 
this satellite peak, the S=3/2 spin state with the d$^{3}$ (Mn$^{4 + })$ 
states are empirically conceived for the Mn 3d spin state \cite{rf:17}. 
Corresponding to the degeneracy of this spin state, the intensity ratio of 
the spin-exchange satellite compared to the main peak is known as 
(2S)/(2S+2) in the atomic model. According to this simplest model, the 
intensity ratio of the satellite will be 3/5=60 {\%} of the main peak for 
the case of S=3/2 spin state. For larger momentum value such as S=2 
(d$^{4}$, Mn$^{3 + })$ or S=5/2 (d$^{5}$, Mn$^{2 + })$ in the high spin 
state, the ratio will become larger. The experimentally derived ratio of 26 
{\%} is much smaller than these values. Thus, further consideration is 
required to interpret the observed results.

From the neutron diffraction measurements, several different spin alignments 
are proposed for the lower temperature antiferromagnetic phases \cite{rf:8,rf:9,rf:10,rf:11,rf:12}.
 Some of them predict complex magnetic structures with 
non-equivalent electronic states in the Mn$_{II}$ 6(g) site \cite{rf:8,rf:9,rf:11}. In 
most cases, smaller magnetic moments are predicted for the Mn$_{I}$ 4(d) 
site (as summarized in Ref. 12). Furthermore, the band calculation predicts 
that the crystallographic Mn$_{I}$ 4(d) site has rather itinerant 3d 
electrons and the Mn$_{II}$ 6(g) site has rather localized ones \cite{rf:1}. In the 
photoemission spectroscopy, spin-exchange interaction effect is commonly 
seen as a localized feature both above and below a magnetic transition 
temperature \cite{rf:14,rf:15}. If we apply these electronic configurations such as 
itinerant and localized Mn 3d electrons to the present case at 195 K in the 
paramagnetic state, it is conceivable that the spin-exchange satellite will 
be mostly derived from the Mn$_{II}$ 6(g) site. Consequently, the Mn 3s 
spectrum can be deconvoluted into three components: namely the larger peak 
at E$_{B}$=82 eV consists of the Mn(1) and Mn(2) components whereas the 
smaller satellite peak at E$_{B}$=86.5 eV consists of one component, Mn(3). 
Then, the intensity of the satellite Mn(3) component comes close to 3/5=60 
{\%} of the Mn(2) intensity in reasonable agreement with the prediction for 
the S=3/2 state. Therefore, we come to the conclusion that the Mn(2) and 
Mn(3) components are the main and the satellite peaks split by the 
spin-exchange interaction. Although the Mn(1) component should also have the 
spin-exchange satellite, the energy separation between them will be small 
enough and the satellite structure may not be clearly recognizable because 
of the FWHM and $\alpha $. Eventually, the intensity of the Mn(1) comes to 
68 {\%} of the `total' intensity of Mn(2)+Mn(3) in agreement with the ratio 
of Mn(1):Mn(2)$\sim $2:3 as in the case of the Mn 2p and 3p spectra.

Table 1 summarizes the fitting parameters for each spectrum; binding energy, 
energy difference between the Mn(1) and Mn(2), asymmetry parameter $\alpha 
$, intensity ratio of Mn(1)/Mn(2), and the experimental full width at half 
maximum (FWHM). Larger FWHM and $\alpha $ are seen for all the Mn(1) 
components compared with the Mn(2) components. The peak asymmetry $\alpha $ 
is related to the infrared divergence or the high density of states (DOS) of 
the Mn 3d states near the Fermi level. The common tendency of higher $\alpha 
$ of the Mn(1) component seen for all 2p, 3p and 3s spectra suggests the 
higher Mn 3d DOS near E$_{F}$ on the Mn(1) site. We further note that all 
$\alpha $ of the Mn core level spectra are more than ten times larger than 
that of the Si 2p state. This suggests that even the Mn(2) sites have some 
\textit{itinerancy} compared to the Si site. The intensity ratios of Mn(1)/Mn(2), especially 
Mn(1)/{\{}Mn(2)+Mn(3){\}} in the case of Mn 3s, are estimated as 0.63$\sim 
$0.68. These values are very consistent with the site occupancy ratio of 2/3 
between the Mn$_{I}$ 4(d) and Mn$_{II}$ 6(g) sites. Together with the 
discussion of the Mn 3s spin-exchange satellite, we conclude that the Mn(1) 
and Mn(2) components in the photoemission spectra are associated with the 
electronic states in the crystallographic Mn$_{I}$ 4(d) and Mn$_{II}$ 6(g) 
sites, respectively. In the Curie-Weiss type paramagnetic phase of 
Mn$_{5}$Si$_{3}$ with metallic conduction above 99 K, it is experimentally 
revealed that the spin is still large on the Mn$_{II}$ 6(g) site, whereas 
the magnitude of the spin is marginal on the Mn$_{I}$ 4(d) site.

\begin{table}[htbp]
\begin{center}
\begin{tabularx}{138mm}{|X|X|X|X|X|X|X|X|}
\hline
\multicolumn{2}{|c|}{orbital} & \multicolumn{1}{c|}{E$_{B}$ (eV)} & \multicolumn{1}{c|}{$\Delta$E$_{B}$ (1)-(2)} & \multicolumn{1}{c|}{$\alpha$} & \multicolumn{1}{c|}{Intensity ratio} & \multicolumn{2}{c|}{FWHM(eV)} \\\cline{7-8}
\multicolumn{2}{|c|}{} & \multicolumn{1}{c|}{} & \multicolumn{1}{c|}{(eV)} & \multicolumn{1}{c|}{} & \multicolumn{1}{c|}{(1)/(2)} & \multicolumn{1}{c|}{j=3/2} & \multicolumn{1}{c|}{j=1/2} \\\hline
\multicolumn{2}{|c|}{Si 2p} & \multicolumn{1}{c|}{99.00} & \multicolumn{1}{c|}{-} & \multicolumn{1}{c|}{0.02} & \multicolumn{1}{c|}{-} & \multicolumn{1}{c|}{0.43} & \multicolumn{1}{c|}{0.42} \\\hline
\multicolumn{1}{|c|}{Mn 2p} & \multicolumn{1}{c|}{(1)} & \multicolumn{1}{c|}{638.89} & \multicolumn{1}{c|}{0.66} & \multicolumn{1}{c|}{0.38} & \multicolumn{1}{c|}{0.66} & \multicolumn{1}{c|}{1.26} & \multicolumn{1}{c|}{1.63} \\\cline{2-3}\cline{5-5}\cline{7-8}
\multicolumn{1}{|c|}{} & \multicolumn{1}{c|}{(2)} & \multicolumn{1}{c|}{638.23} & \multicolumn{1}{c|}{} & \multicolumn{1}{c|}{0.31} & \multicolumn{1}{c|}{} & \multicolumn{1}{c|}{0.59} & \multicolumn{1}{c|}{1.32} \\\hline
\multicolumn{1}{|c|}{Mn 3p} & \multicolumn{1}{c|}{(1)} & \multicolumn{1}{c|}{47.86} & \multicolumn{1}{c|}{0.98} & \multicolumn{1}{c|}{0.33} & \multicolumn{1}{c|}{0.63} & \multicolumn{1}{c|}{1.77} & \multicolumn{1}{c|}{2.23} \\\cline{2-3}\cline{5-5}\cline{7-8}
\multicolumn{1}{|c|}{} & \multicolumn{1}{c|}{(2)} & \multicolumn{1}{c|}{46.88} & \multicolumn{1}{c|}{} & \multicolumn{1}{c|}{0.26} & \multicolumn{1}{c|}{} & \multicolumn{1}{c|}{1.16} & \multicolumn{1}{c|}{1.21} \\\hline
\multicolumn{1}{|c|}{Mn 3s} & \multicolumn{1}{c|}{(1)} & \multicolumn{1}{c|}{82.17} & \multicolumn{1}{c|}{0.53} & \multicolumn{1}{c|}{0.36} & \multicolumn{1}{c|}{(1)/\{(2)+(3)\}} & \multicolumn{2}{c|}{2.00} \\\cline{2-3}\cline{5-8}
\multicolumn{1}{|c|}{} & \multicolumn{1}{c|}{(2)} & \multicolumn{1}{c|}{81.64} & \multicolumn{1}{c|}{} & \multicolumn{1}{c|}{0.25} & \multicolumn{1}{c|}{0.68} & \multicolumn{2}{c|}{1.95} \\\cline{2-8}
\multicolumn{1}{|c|}{} & \multicolumn{1}{c|}{(3)} & \multicolumn{1}{c|}{86.19} & \multicolumn{1}{c|}{(3)-(2)} & \multicolumn{1}{c|}{0.28} & \multicolumn{1}{c|}{(3)/(2)} & \multicolumn{2}{c|}{3.03} \\\cline{4-4}\cline{6-6}
\multicolumn{1}{|c|}{} & \multicolumn{1}{c|}{} & \multicolumn{1}{c|}{} & \multicolumn{1}{c|}{4.55} & \multicolumn{1}{c|}{} & \multicolumn{1}{c|}{0.58} & \multicolumn{2}{c|}{} \\\hline
\end{tabularx}
\end{center}
\caption{Fitting parameters used for the deconvolution of the inner core spectra; binding energies, energy shifts, asymmetry parameter $\alpha$, intensity ratios of Mn(1)/Mn(2), and experimental full width at half maximum (FWHM). For the Mn 3s spectrum are given the energy shifts of Mn(3) from Mn(2) components and the intensity ratios of Mn(1)/\{Mn(2)+Mn(3)\} and Mn(3)/Mn(2).}
\label{table:1}
\end{table}%

Figure 2 shows the valence band spectrum measured at h$\nu $=900 eV. A 
secondary electron background is subtracted from the experimental spectrum 
in the figure. Three peak structures are recognized in the experimental 
spectrum up to E$_{B}$=7 eV. The cross section of the Si 2p states is ten 
times smaller than that of the Mn 3d states at 900 eV \cite{rf:18}. Therefore, the 
photoemission features in the region of E$_{B}$=3.0 eV$\sim $E$_{F}$ are 
mostly derived from the Mn 3d states. In addition, the measurement at this 
h$\nu$ is rather bulk-sensitive compared with the lower h$\nu$ 
photoemission \cite{rf:19,rf:20}. The present spectrum is therefore reflecting the DOS 
of bulk Mn 3d states.

\begin{figure}
\begin{center}
\includegraphics[width=7cm,clip]{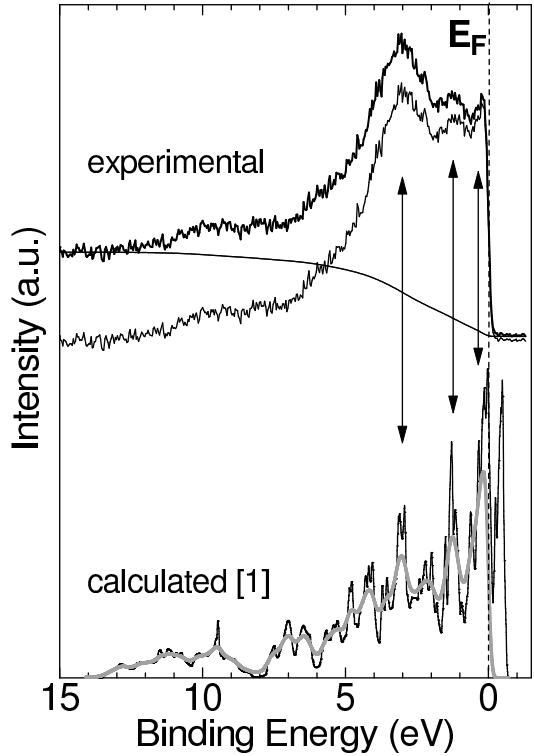}
\end{center}
\caption{High-resolution photoemission spectrum of valence band at 195 K for h$\nu$=900 eV compared with the band calculation for the paramagnetic state of Mn$_5$Si$_3$ done by L. Vinokurova et al. \cite{rf:1}}
\label{fig:2}
\end{figure}%

The total DOS predicted by the band calculation by L. Vinokurova et al. is 
shown by the thin solid curve in the lower panel \cite{rf:1}. For comparison, the 
thick solid curve shows its tentatively broaden spectrum with considering 
the Lorentzian FWHM of 200 meV, the experimental energy resolution of 100 
meV, and the Fermi-Dirac function at 195 K. In the experimental spectrum, 
one observes a sharp peak near the Fermi level and broader peaks at the 
binding energies of 1.3 and 3.0 eV. They correspond roughly to the peaks of 
DOS of the band calculation as connected by arrows. The observed hump near 
10 eV corresponds to the DOS of the Si 3s state which is commonly seen in 
photoemission spectra of 3d transition metal silicides \cite{rf:21}. In this way, 
the valence band bulk photoemission spectrum of Mn$_{5}$Si$_{3}$ agrees 
qualitatively with the result of the band calculation.

\section{Conclusion}
We have performed the bulk-sensitive high-resolution photoemission 
measurements on a single-crystal Mn$_{5}$Si$_{3}$ by using synchrotron 
radiation. Each Mn core-level spectrum is found to consist of two components 
associated with crystallographic sites of Mn$_{I}$ 4(d) and Mn$_{II}$ 6(g). 
The large asymmetry $\alpha $ of the photoemission spectra of the Mn core 
states indicates that the DOS near the Fermi energy consists mostly of the 
Mn 3d electron states. The valence band spectrum corresponds well to the DOS 
of the bulk electronic states.

\section{Acknowledgements}
We are grateful to Dr. T. Muro and Dr. Y. Saitoh for their assistance as 
beam line scientists. The research was performed at SPring-8 under the 
support of a Grant-in-Aid for the COE Research (10CE2004) of the Ministry of 
Education, Culture, Sports, Science, and Technology (MEXT), Japan.



\end{document}